\documentclass[twoside]{hs04}
\def\runauthor{M\'onica Luisa V\'azquez Acosta}
\def\shorttitle{Exotic hadronic states at HERA}
\begin{document}

\title{EXOTIC HADRONIC STATES AT HERA}

\author{M\'onica V\'azquez}

\address{
ON BEHALF OF THE ZEUS, H1, HERMES and HERA-B COLLABORATIONS\\
National Institute for Nuclear Physics and High Energy Physics NIKHEF\\
Kruislaan 409, 1098 SJ Amsterdam, The Netherlands\\
E-mail: monicava@mail.desy.de }

\maketitle

\abstracts{
Recent results from H1, ZEUS, HERMES and HERA-B on searches for exotic baryons in ep collisions, eD scattering and pA scattering at HERA are reviewed. Evidence for the production of the strange pentaquark $\Theta^{+}$ and of a narrow anti-charmed baryon decaying to $D^{*-}p$ together with negative results of pentaquark searches at HERA are presented.
}

\section{Introduction}

The hadronic states studied in hadron spectroscopy have been successfully explained by quark-antiquark states (mesons) or three quark states (baryons). States whose quantum numbers could be understood in terms of  more complex quark structures were up to now not observed. Such exotic states were already proposed on the basis of quark and bag models~\cite{RLJ} in the early days of QCD, with the hope that the discovery of such states would provide new insights into the dynamics of quark interactions. The Skyrme model~\cite{AM,MC} predicts new exotic states belonging to higher SU(3) representations. Using this model, Praszalowicz~\cite{MP} provided the 
first estimate of the lightest exotic state of M$\sim$1530 MeV.  The Chiral Quark Soliton model~\cite{MC} was aused to obtain an exotic baryon of spin 1/2, isospin 0 and strangeness S=+1. In this approach~\cite{HW,DPP} the baryons are rotational states of the soliton nucleon in spin and 
isospin space, and the lightest exotic baryon lies at the apex of an anti-decuplet with spin 1/2, 
which corresponds to the third rotational excitation in a three flavour system. Treating the known 
N(1710) resonance as a member of the antidecuplet, Diakonov, Petrov and Polyakov~\cite{DPP} derived a 
mass of 1530 MeV and a width of less than 15 MeV for this exotic baryon,  named the 
$\Theta^{+}$. It corresponds to a $uudd\bar{s}$ configuration, and decays through the channels 
$\Theta^{+}\rightarrow pK^0_S$ or $nK^+$. 

Experimental evidence for an exotic baryon first came recently~\cite{TN} from the observation of a narrow resonance at $1540 \pm 10$ MeV in the $K^-$ missing mass spectrum for the $\gamma n \rightarrow K^+K^-n$ reaction on $^{12}C$. The decay mode corresponds to a $S=+1$ resonance and signals which can be associated with an exotic pentaquark state with content $uudd\bar{s}$. Confirmation came quickly from a series of experiments, with the observation of sharp peaks~\cite{DIANA,CLAS,SAPHIR,AEA} in the $nK^{+}$ and $pK_S^{0}$ invariant mass spectrum near 1540 MeV, in each case with a width limited by the experimental resolution. The failure to observe a corresponding $\Theta^{++}$ peak in the $pK^+$ invariant mass spectrum in some of these experiments was taken to suggest that the state is an isospin singlet. 

The $\Theta^{+}$ has been observed both in fixed target experiments and high energy experiments. In the case of fixed target experiments the $\Theta^{+}$ can originate from the valence quarks as opposed to high energy experiments were the $\Theta^{+}$ is produced in the fragmentation.

The baryon states at the bottom two vertices of the anti-decuplet are also exotic. Strong evidence in support of the baryon decuplet comes from the reported observation of an exotic $S=-2$, $Q=-2$ baryon resonance in proton-proton collisions at $\sqrt{s}=17.2$ GeV at the CERN SPS~\cite{NA49}. A narrow peak at a mass of about 1862 MeV in the $\Xi^{-}\pi^{-}$ invariant mass spectrum is proposed as a candidate for the predicted exotic $\Xi^{--}_{\frac{3}{2}}$ baryon with $S=-2$, $I=\frac{3}{2}$ and a quark content of $dsds\bar{u}$. At the same mass, 
a peak is observed that is a candidate for $\Xi^{0}_{\frac{3}{2}}$. The corresponding anti-baryon spectra shows enhancement at the same mass. 

This paper presents the results of the searches for strange pentaquarks from HERMES~\cite{HERMES}, ZEUS~\cite{ZEUSTHETA,ZEUSTHETA2,ZEUSCASCADE} and HERA-B~\cite{HERAB}, and for an anti-charmed baryon decaying into $D^{*-}p$ from H1~\cite{H1} and ZEUS~\cite{ZEUSDSTAR}.

\section{Kinematics at HERA}
HERA is a positron and proton storage ring with four experiment halls. 
Positrons with an energy of 27.5 GeV are collided on protons of 820-920 GeV in two interaction regions (H1 and ZEUS) yielding a centre-of-mass energy of $\sqrt{s} =300 - 318$ GeV. In a third interaction region (HERMES) the positrons interact on a deuteron target at $\sqrt{s} =7.2$ GeV. In the last interaction region (HERA-B) the protons interact on a carbon (C), titanium (Ti) or tungsten (W) target at $\sqrt{s} = 41.6$ GeV.

The kinematics of the lepton-nucleon scattering is described by three independent variables: the centre-of-mass energy $\sqrt{s}$, the four-momentum transfer squared $q^2 = -Q^2$ and either the scaling variable $x=Q^2/2P\cdot q$ or the inelasticity $y=P\cdot q/P\cdot k$, where $P$ and $k$ denote the four-momentum of the nucleon and the lepton, respectively. The $\gamma p$ centre-of-mass energy squared is given by $W_{\gamma p}^2 \approx y\cdot s - Q^2$.

\begin{figure} [t]
\vspace*{-.2cm}
  \unitlength 1cm
\begin{minipage}{6cm}
\vspace*{0.2cm}
  \begin{picture}(6,6)
    \put(0.,0.){\psfig{file=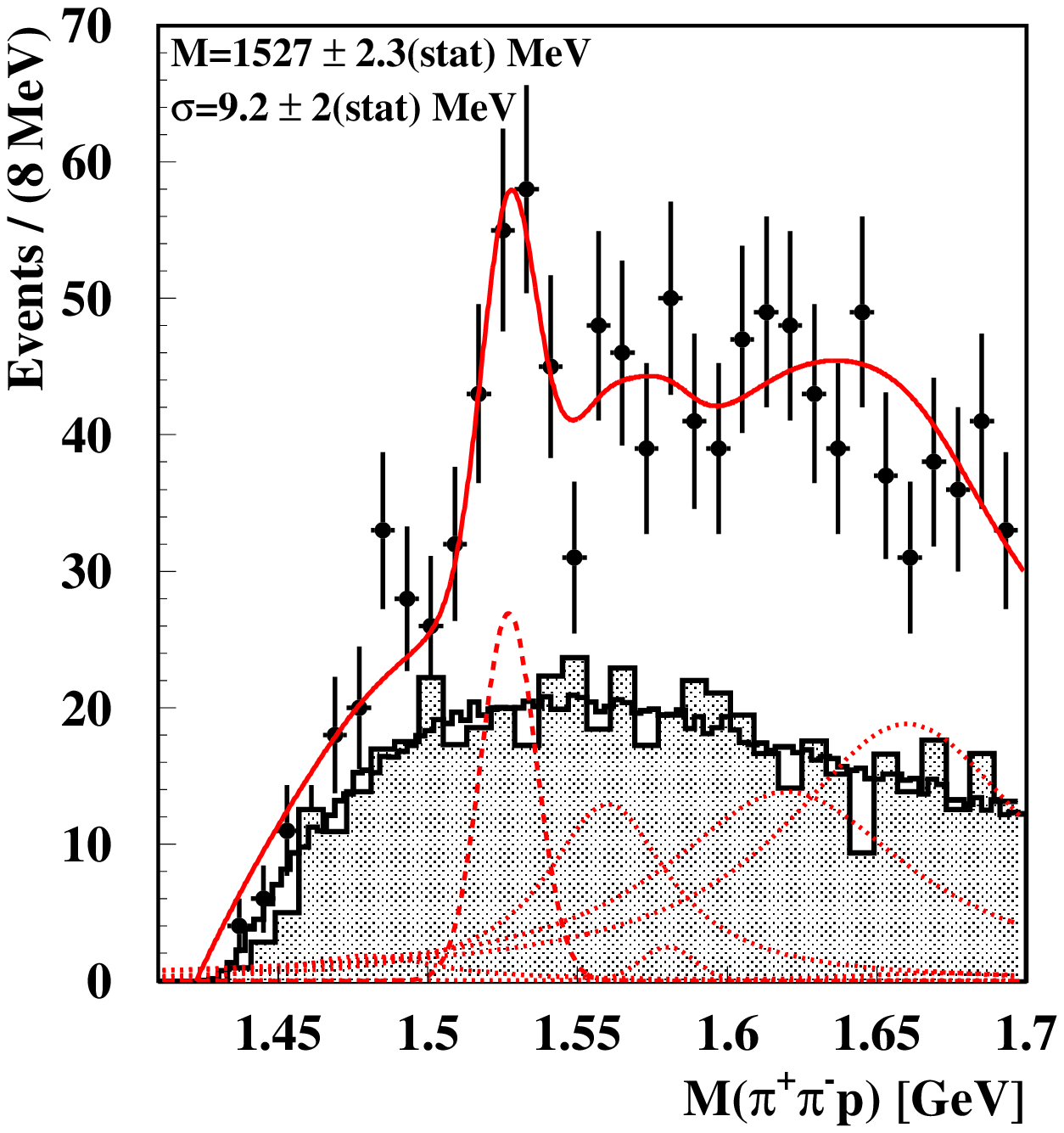,height=6cm,clip=}}    
    \put(4.5,5.){\tiny(a)}
  \end{picture}
\end{minipage}
\hfill
\begin{minipage}{6cm}
\begin{picture}(6,6)
    \put(0.,0.){\psfig{file=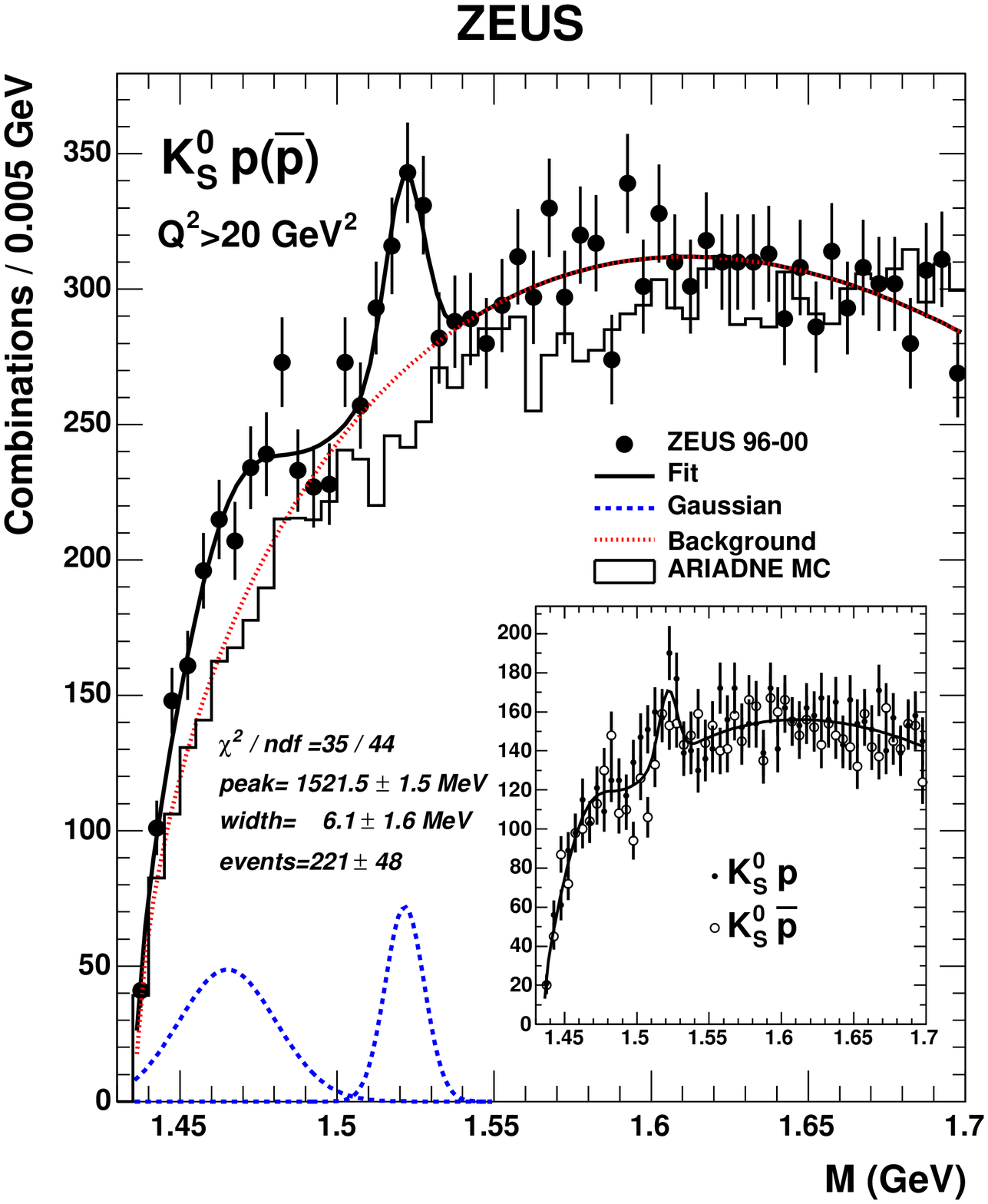,height=6cm,clip=}}
    \put(4.3,5.){\tiny(b)}
  \end{picture}
\end{minipage}
\caption{
Invariant $M_{p\pi^+\pi^-}$ mass distribution observed by (a) HERMES at $Q^2>1$ GeV$^2$ and (b) ZEUS at $Q^2>20$ GeV$^2$.
\label{plot_Theta}
}
\end{figure}

\section{Search for strange pentaquarks}
The production of $\Theta^{+}$ has been studied via its decay into $K^0_Sp$ in three  different kinematic regions by HERMES~\cite{HERMES}, ZEUS~\cite{ZEUSTHETA,ZEUSTHETA2} and HERA-B~\cite{HERAB}. Furthemore the ZEUS~\cite{ZEUSCASCADE} and the HERA-B ~\cite{HERAB} collaborations have searched for the S=-2 baryons $\Xi^{--}_{\frac{3}{2}}$ and $\Xi^{0}_{\frac{3}{2}}$.
\subsection{Search for $\Theta^+$}
The HERMES collaboration has performed a $\Theta^+$ search using the decay chain $\Theta^+ \rightarrow p K_S^{0} \rightarrow p \pi^+ \pi^-$. The sample used is eN scattering data with a longitudinally polarized deuterium gas target having an integrated luminosity of 250 pb$^{-1}$. The yields were summed over two spin orientations. The kinematic region is restricted to $0.02 < x < 0.8$, $Q^2 > 1$ GeV$^2$ and $W>2$ GeV. Hadron identification is accomplished with the Ring-Imaging Cerenkov detector which provides good separation of pions, kaons and protons. Identified protons are combined with well identified secondary vertices with an invariant $M_{\pi^+\pi^-}$ mass within $2\sigma$ of the reconstructed $K_S^{0}$ mass. Possible $\Lambda$ contamination is suppressed by rejecting  $K_S^{0}$ candidates with a $M_{p\pi^-}$ mass within $2\sigma$ of the nominal $\Lambda$ mass.

The resulting $M_{p{\pi}^+\pi^-}$ mass distribution is shown in Figure~\ref{plot_Theta}a. A narrow peak structure is observed around the $\Theta^+$ mass. No such structure is obtained when the $\pi^+\pi^-$ mass combinations from the $K_S^{0}$ side bands are used instead. the data are also compared with expectations from the PYTHIA6 Monte Carlo simulation~\cite{PYT} (gray shaded histogram) and from the mixed-event model (fine-binned histogram) normalized to PYTHIA6. In the second model,  mixed-event model, it is assumed that the 4-momenta of the $K_S^{0}$ and the proton are largely uncorrelated. The background can then be simulated by combining a kaon and a proton from different events which satisfy the same cuts as in the original analysis. 
No peak structure is visible in the Monte Carlo or the mixed-event model expectations. The PDG~\cite{PDG} reports the possible existance of several $\Sigma$ bumps decaying to $N\overline{K}$ in this mass region. These are not included in the simulation and may account for the discrepancies. 

The fit to the data shown in Fig.~\ref{plot_Theta}a (smooth solid line), which is based on the mixed-event model, the $\Sigma$ bumps  (dotted curves) and a Gaussian (dashed curve) for a possible $\Theta^+$ signal, yields a good description. A peak of about 80 $\Theta^+$ events with a significance of $4.3\sigma$ is observed at a mass of $M=1527\pm 2.3 (stat.)$ MeV. 

A similar analysis has been performed in the ZEUS collaboration at higher energies using the $ep$ data taken in the years 1996-2000 with an integrated luminosity of $121$ pb$^{-1}$. The kinematic region is restricted to $Q^2>1$ GeV$^2$ and $0.01 \le y \le 0.95$.

The decay chain $\Theta^+ \rightarrow p K_S^{0} \rightarrow p \pi^+ \pi^-$ has also been used. About 866800 $K_S^{0}$ candidates are selected for $Q^2>1$ GeV$^2$. They are combined with proton candidates selected via the energy-loss measurement, $dE/dx$, in the central tracking chamber.

The $M_{p{\pi}^+\pi^-}$ mass distribution shows sign of structure below about 1600 MeV. For $Q^2>10$ GeV$^2$, a peak is seen in the mass distribution around 1520 MeV. In Figure~\ref{plot_Theta}b the $M_{p{\pi}^+\pi^-}$ is shown for $Q^2>20$ GeV$^2$. The figure includes the Monte Carlo expectation from ARIADNE~\cite{ARI} scaled to the data for $M_{p{\pi}^+\pi^-} > 1650$~MeV. After scaling ARIADNE does not describe the data at low masses, maybe due to the absence of the $\Sigma$ bumps in the simulation. 

A fit to the data of a smooth background function and two Gaussians, also shown in Figure~\ref{plot_Theta}b, gives a signal of $221 \pm 48$ events at a mass of $1521.5\pm 1.5(stat.)$~MeV with a significance of $4.6 \sigma$. The Gaussian width of $6.1$ MeV is found to be consistent with the experimental resolution. The signal is observed at similar rate for protons and for antiprotons suggesting the existance of the anti-pentaquark $\Theta^-$.

\begin{figure} [t]
  \unitlength 1cm
\begin{minipage}{6cm}
  \begin{picture}(6,6)
    \put(0.,0.){\psfig{file=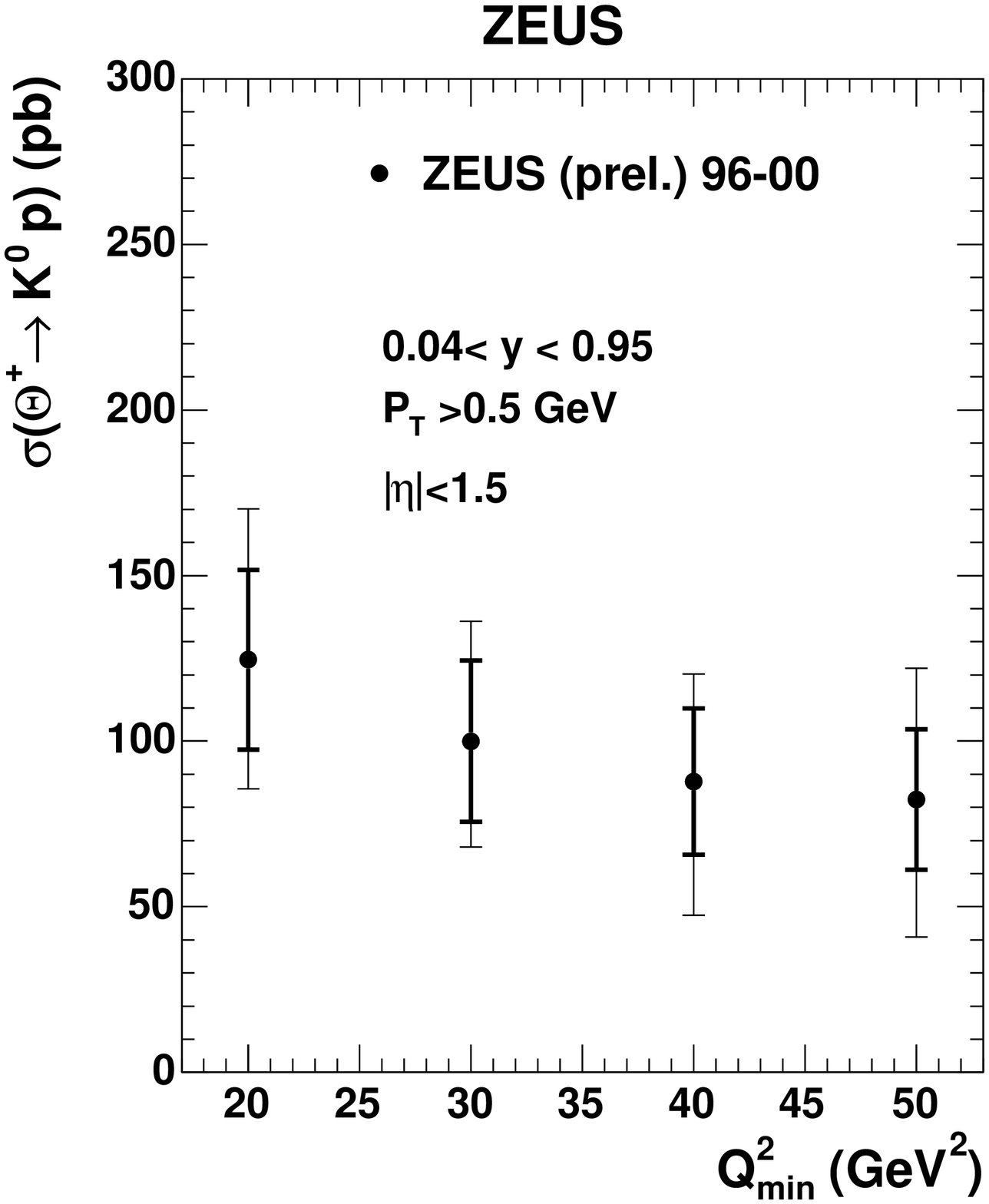,height=6.0cm,clip=}}    
    \put(1.,1.){\tiny(a)}
  \end{picture}
\end{minipage}
\hfill
\begin{minipage}{6cm}
\begin{picture}(6,6)
    \put(0.,0.){\psfig{file=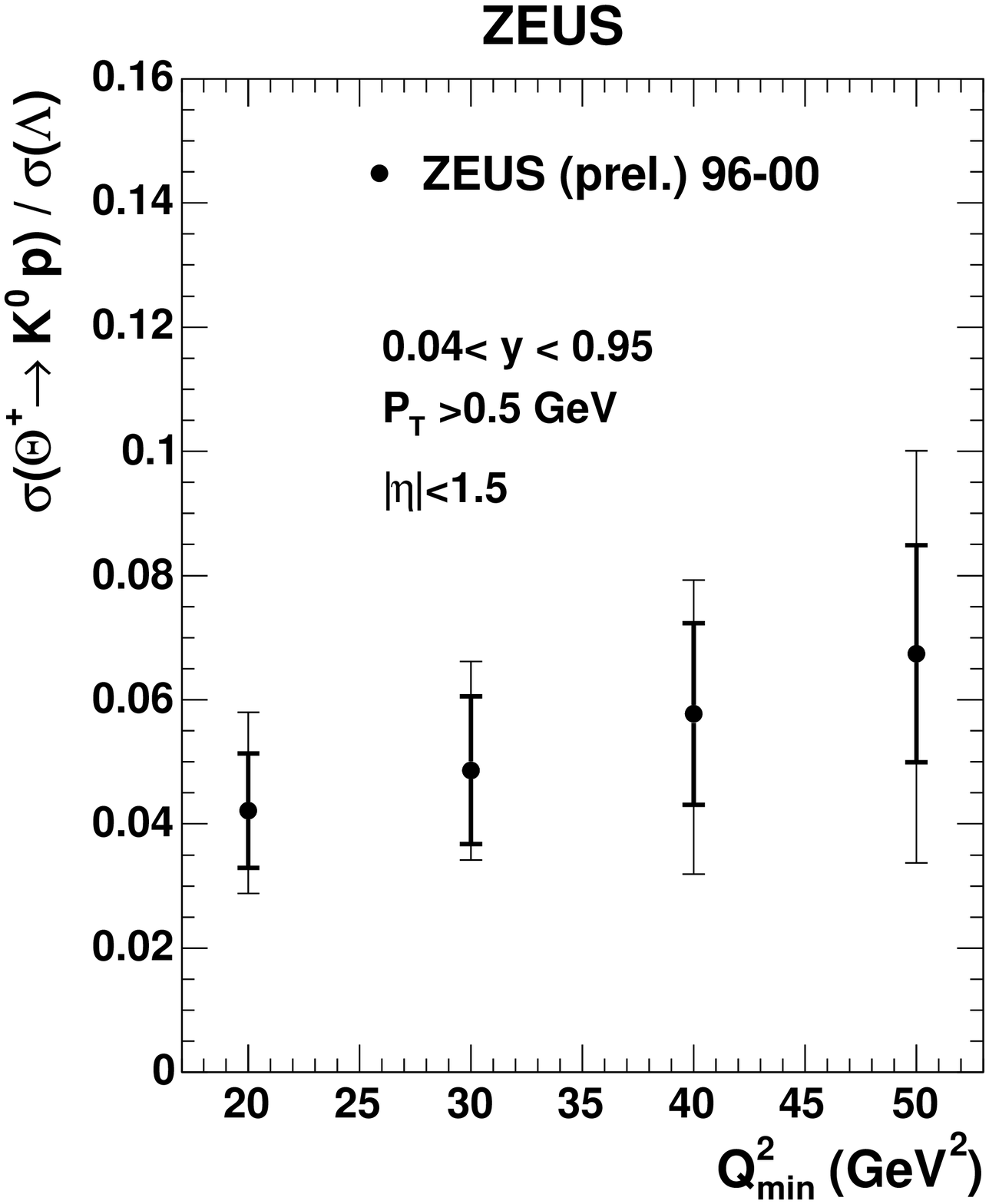,height=6.0cm,clip=}}
    \put(1.,1.){\tiny(b)}
  \end{picture}
\end{minipage}
\caption{ZEUS measurements of 
(a) cross sections for the 1522 MeV baryonic state decaying to $K^0_S p(\bar{p})$ integrated above $Q^2_{min}$, (b) the ratio of the cross section $\sigma(\Theta^+\rightarrow K^0_S p(\bar{p}))$ to the inclusive $\Lambda$ cross section, $\sigma(\Lambda+\bar{\Lambda})$,integrated above $Q^2_{min}$.
\label{plot_zeusThetacross}
}
\end{figure}

The ZEUS collaboration has also measured the cross section for the production of the $\Theta^+$ baryons and their antiparticles in the kinematic region $Q^2>20$ GeV$^2$, $0.04 \le y \le 0.95$, $p_T > 0.5$ GeV and $|\eta|< 1.5$,
\begin{displaymath}
\sigma (e^\pm p \rightarrow e^\pm \Theta^+ X \rightarrow e^\pm K_S^{0} p X) = 125 \pm 27 (stat.) ^{+36}_{-28} (syst.)\; pb.
\end{displaymath}
Figure~\ref{plot_zeusThetacross}a shows the cross section integrated above $Q^2_{\rm min}$. Figure~\ref{plot_zeusThetacross}b shows the ratio of this cross section to the $\Lambda$ cross section integrated above $Q^2_{\rm min}$, where the ratio, defined in the same kinematic region as above, is
\begin{displaymath}
ratio = \frac{\sigma (e^\pm p \rightarrow e^\pm \Theta^+ X \rightarrow e^\pm K_S^{0} p X)}{\sigma (e^\pm p \rightarrow e^\pm \Lambda X)}.
\end{displaymath}
This ratio for $Q^2_{\rm min}=20$ GeV$^2$, is $4.2\pm 0.9 (stat.) ^{+1.2}_{-0.9}(syst.) \%$ and, in the analyzed data, shows no significant dependence on $Q^2_{\rm min}$. Since the $\Theta^+$ has other decay channels in addition to  $\Theta^+\rightarrow K_S^{0} p$, this ratio sets a lower limit on the production rate of the $\Theta^+$ to that of the $\Lambda$-baryon. 

The HERA-B Collaboration has searched for the $\Theta^{+}$ pentaquark candidates in proton-induced reactions on C, Ti and W targets at mid-rapidity and $\sqrt{s}=41.6$ GeV, in $2 \cdot 10^8$ inelastic events. No evidence for a narrow signal in the $K_S^{0} p$ spectrum is found.  The 95\% confidence level (C.L.) upper limits for the inclusive production cross section times the branching fraction is ${\cal B} d\sigma /dy | _{y \sim 0}$ is 3.7 $\mu b/nucleon$ for a mass of 1530 MeV and 22 $\mu b/nucleon$ for a mass of 1540 MeV.  The upper limit of the ratio $\Theta^+/\Lambda (1520)<2.7\%$ is significantly lower than the model predictions based on the Gribov-Regge approach for describing the $\Theta^{+}$ production and its $\sqrt{s}$ dependence in pp collisions~\cite{GRI}. 

\begin{figure} [t]
  \unitlength 1cm
\begin{minipage}{6cm}
  \begin{picture}(6,6)
    \put(0.,0.){\psfig{file=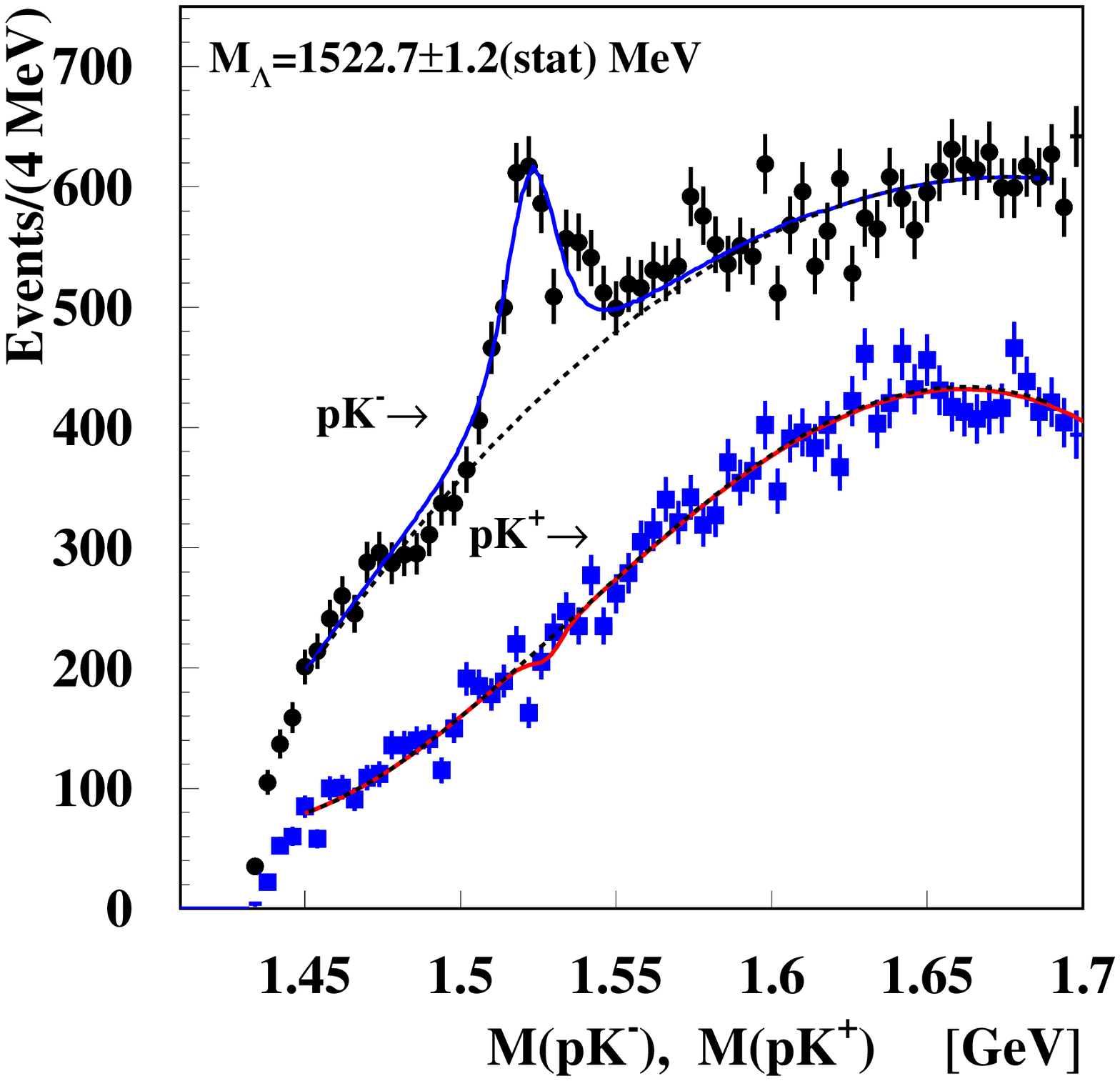,height=6.0cm,clip=}}    
    \put(5.2,5.5){\tiny(a)}
  \end{picture}
\end{minipage}
\hfill
\begin{minipage}{6cm}
\vspace*{-1.5cm}
\begin{picture}(6,6)
    \put(0.,0.){\psfig{file=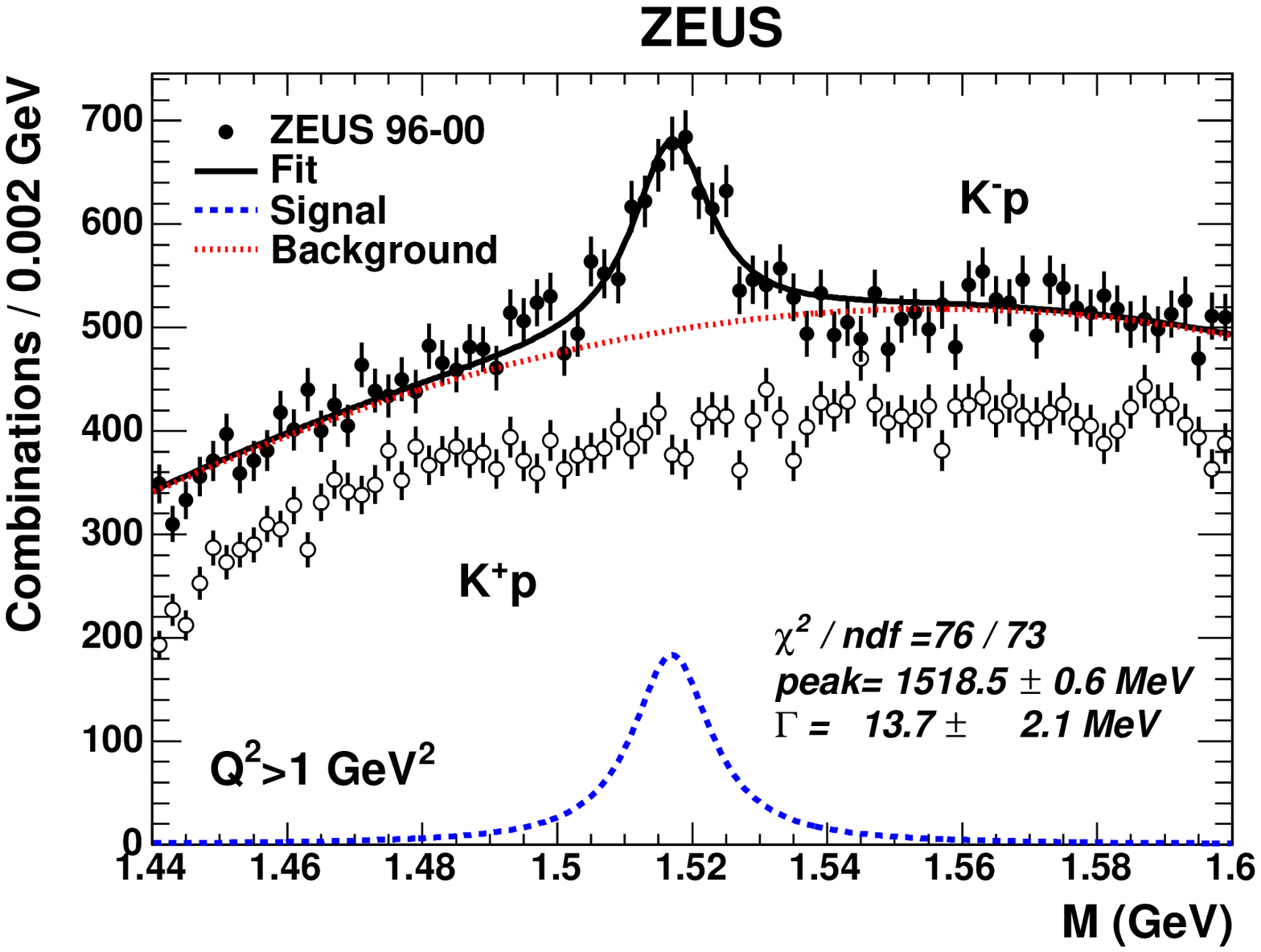,height=4.5cm,clip=}}
    \put(5.2,3.5){\tiny(b)}
  \end{picture}
\end{minipage}
\caption{
Invariant mass distribution $M_{pK^-}$ (top) and $M_{pK^+}$ (bottom) observed by (a) HERMES  and (b) ZEUS.
\label{plot_Thetaplusplus}
}
\end{figure}

HERMES and ZEUS have also searched for the $\Theta^{++}$ signal via its possible decay $\Theta^{++} \rightarrow K^+ \pi^+$. Figure~\ref{plot_Thetaplusplus} show the $M_{pK^-}$  and $M_{pK^+}$ mass spectrum observed by HERMES (Fig.~\ref{plot_Thetaplusplus}a) and ZEUS (Fig.~\ref{plot_Thetaplusplus}b). No peak structure is observed in the $M_{pK^+}$ spectrum but in the $M_{pK^-}$ spectrum the well established resonance $\Lambda (1520) \rightarrow pK^-$ is clearly seen. As no signals are found in the $\Theta^+$ mass range, this suggests that the $\Theta^+$ could be isoscalar.

\begin{figure} [t]
  \unitlength 1cm
\begin{minipage}{6cm}
\vspace*{-1.cm}
  \begin{picture}(6,6)
    \put(0.,0.){\psfig{file=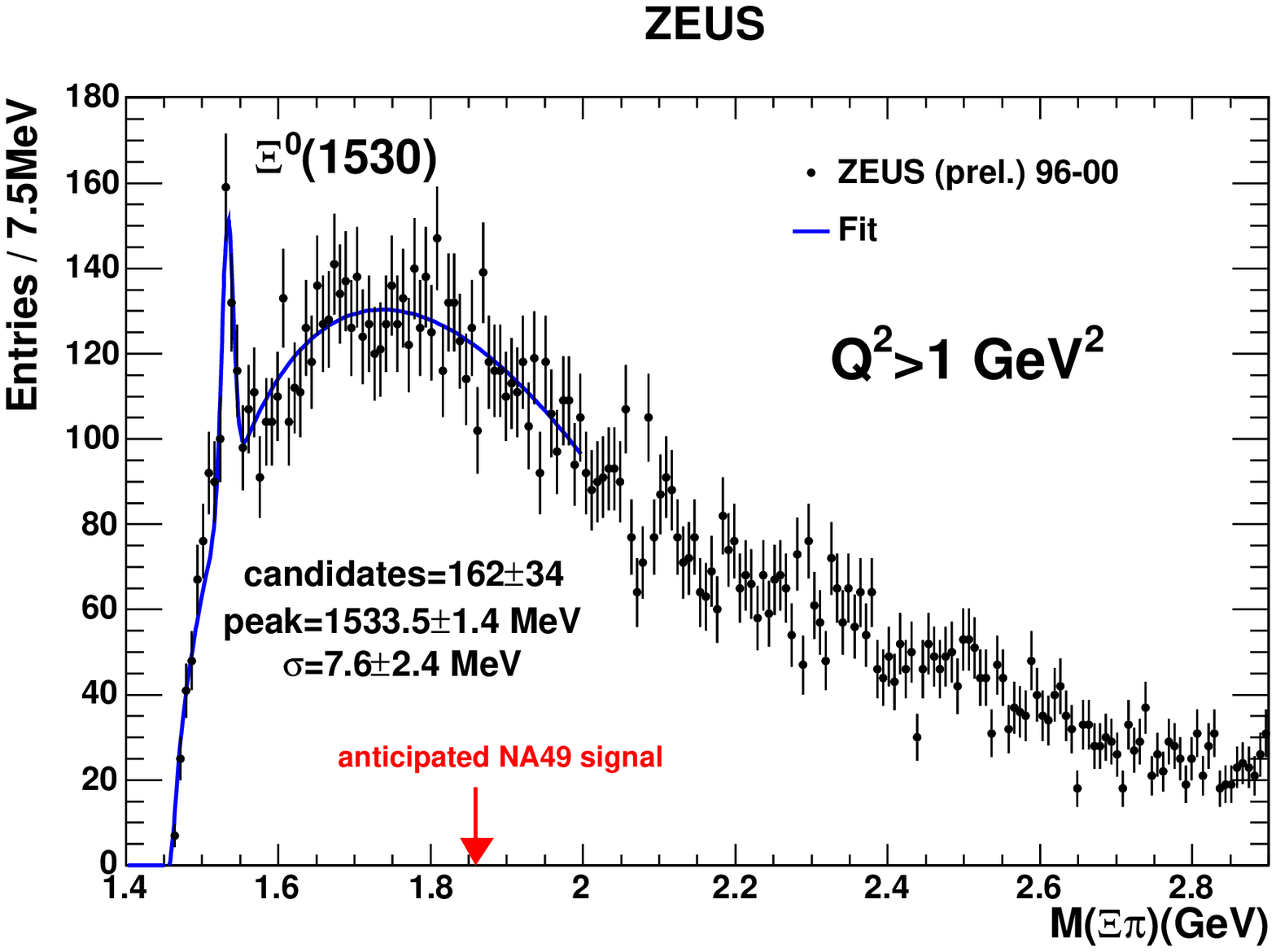,height=4.5cm,clip=}}    
    \put(5.4,3.6){\tiny(a)}
  \end{picture}
\end{minipage}
\hfill
\begin{minipage}{6cm}
\begin{picture}(6,6)
    \put(0.,0.){\psfig{file=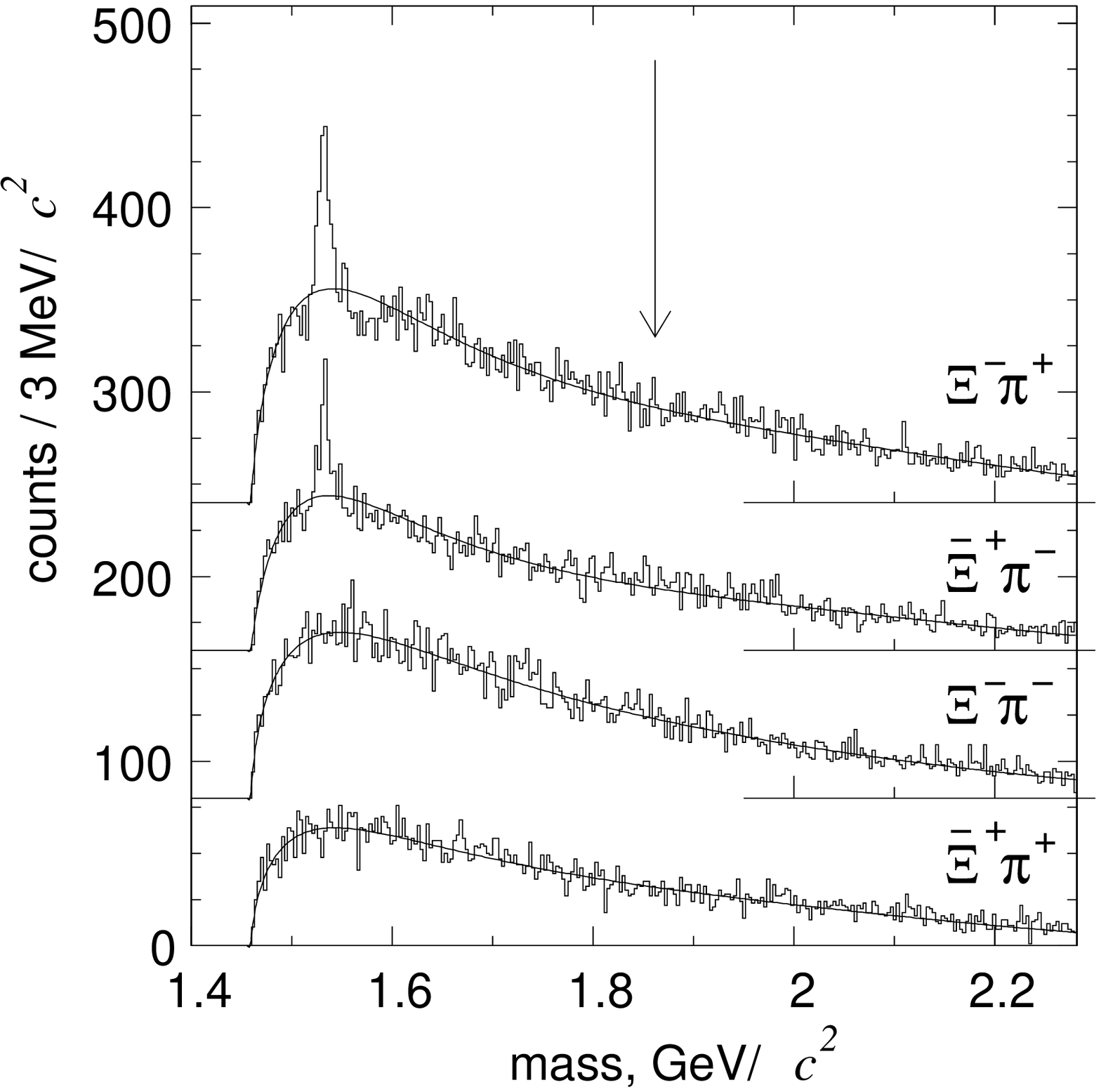,height=5.8cm,clip=}}
    \put(5.2,5.3){\tiny(b)}
  \end{picture}
\end{minipage}
\caption{
Invariant mass distribution $M_{\Xi\pi}$ observed by (a) ZEUS for $Q^2>1$ GeV$^2$ and for all four charge combinations combined (b) HERA-B in p+C collisions  separated in different charge combinations.
\label{plot_cascade}
}
\end{figure}

\subsection{Search for $\Xi^{--}_{\frac{3}{2}}$ and $\Xi^{0}_{\frac{3}{2}}$}
ZEUS has performed an analysis in the channel $\Xi^-\pi^\pm$ to search for the strange pentaquark $\Xi^{--}$ and its neutral partner. The decay chain $\Xi^{--} \rightarrow \Xi^-\pi^- \rightarrow \Lambda\pi^- \pi^-$ has been considered. $\Lambda$ baryons were identified by the charged-decay mode, $\Lambda \rightarrow p \pi^-$ , using pairs of tracks from secondary vertices. 
 These are then combined with another pion from the primary vertex. Figure~\ref{plot_cascade}a shows the $M_{\Xi\pi}$ mass distribution for all possible $\Xi\pi$ charge combinations for $Q^2>1$ GeV$^2$. While the $\Xi^0(1530)$ is clearly visible, no signal is observed around the 1860 MeV as observed by the NA49 collaboration~\cite{NA49}. Even when restricting to $Q^2>20$ GeV$^2$, where the $\Theta^+$ signal was best seen by ZEUS, no signal is observed.

HERA-B has also searched for the strange pentaquark $\Xi^{--}$ in the $\Xi^-\pi^\pm$ channels in proton-induced reactions on C, Ti and W targets. 
In the analysis,  clean signals for $\Xi^-\rightarrow\Lambda\pi^-$ are obtained by requesting the $\Lambda\pi^-$ vertex to be at least 2.5 cm downstream of the target and the event to exhibit a cascade topology: a further downstream $\Lambda$ vertex and a $\Xi^-$ pointing back to the target wire (impact parameter $b < 1$ mm). The pion candidates were required to originate from the primary vertex. Figure~\ref{plot_cascade}b shows the $M_{\Xi\pi}$ mass distribution for the C target, separated in the different charge combinations. In the neutral channels the $\Xi^{0}(1530)$ resonance is seen with a signal of $\sim 10^3$ events. The observed width of ($\sim 9.5$ MeV) agrees with the MC simulation. None of the mass spectra of Fig.~\ref{plot_cascade}b show evidence for the narrow resonance reported by the NA49 collaboration.

\section{Search for a narrow charmed baryonic state}
The production of a charmed pentaquark $\Theta_c$ has been studied via its decay into $D^* p$ by H1~\cite{H1} and ZEUS~\cite{ZEUSDSTAR}.

The analysis of H1 is based on the DIS data taken in the years 1996-2000 with a luminosity of $75$ pb$^{-1}$ in the kinematic region $1\le Q^2 \le 100$ GeV$^2$ and $0.05 \le y \le 0.7$. The $D^{*\pm}$ charmed meson has been reconstructed via its decay chain $D^{*+} \rightarrow D^0 \pi_S^{+}  \rightarrow (K^- \pi^+) \pi_S^{+}$. Around 3400 $D^*$ candidates are selected. $D^*$ candidates having a mass difference $\Delta M_{D^*} = m(K\pi\pi_S) - m(K\pi)$ within 2.5 MeV around the nominal $M(D^*) -M(D^0)$ mass difference are combined with proton candidates selected via $dE/dx$.

\begin{figure} [t]
  \unitlength 1cm
\vspace*{-1.5cm}
\begin{minipage}{6cm}
  \begin{picture}(6,6)
    \put(0.,0.){\psfig{file=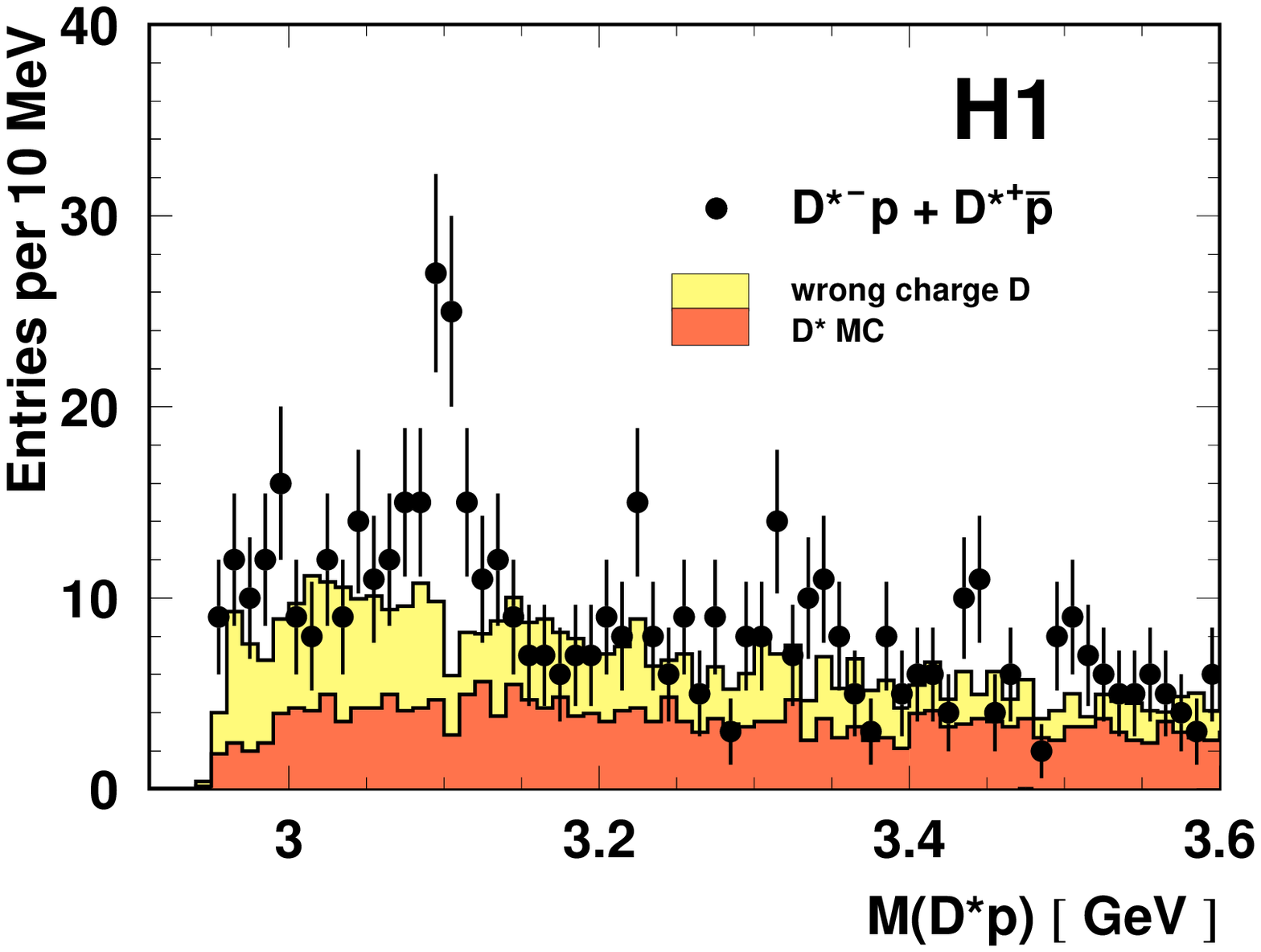,height=4.5cm,clip=}}    
    \put(0.8,4.){\tiny(a)}
  \end{picture}
\end{minipage}
\hfill
\begin{minipage}{6cm}
\begin{picture}(6,6)
    \put(0.,0.){\psfig{file=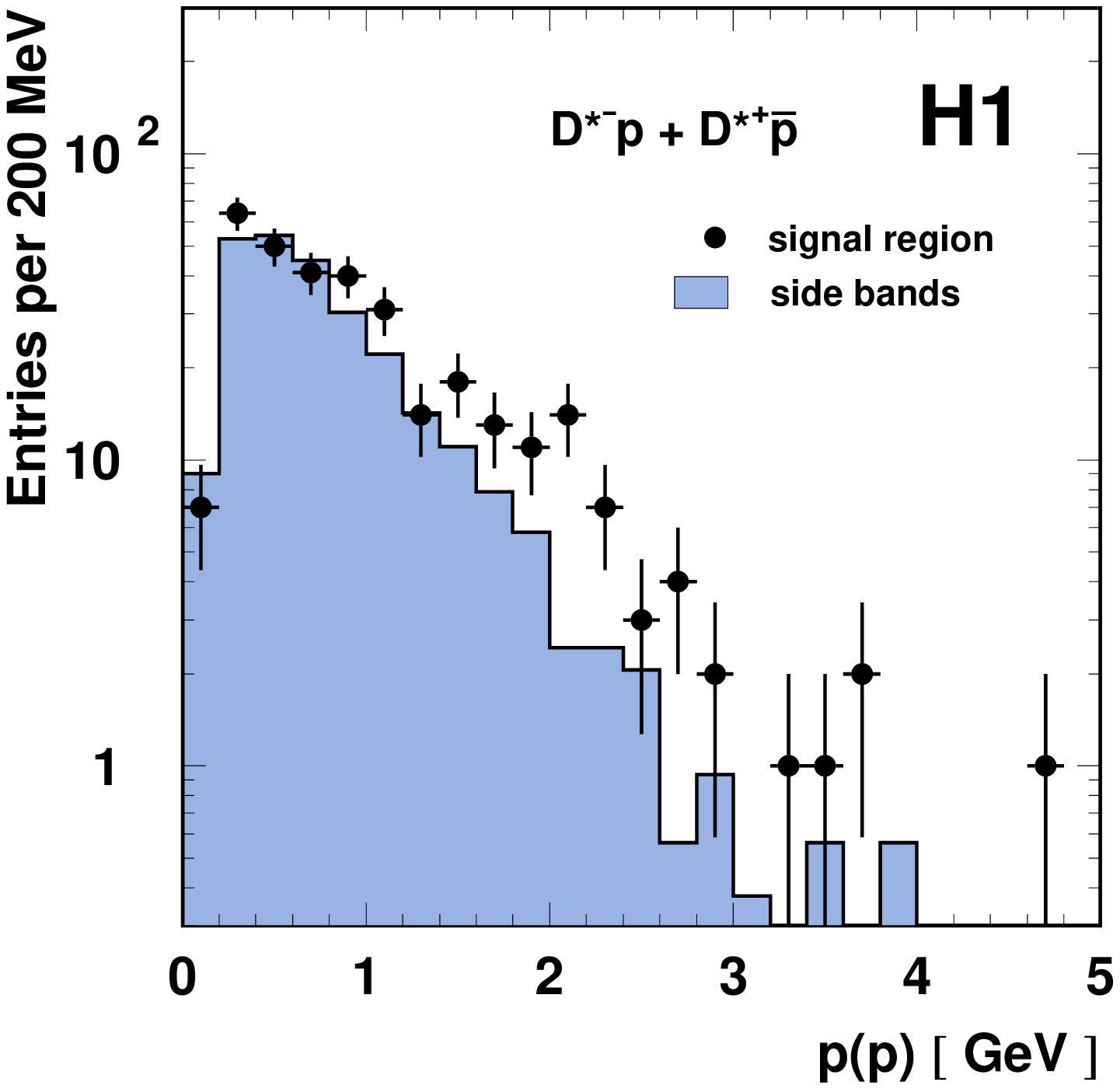,height=4.5cm,clip=}}
    \put(0.8,4.){\tiny(b)}
  \end{picture}
\end{minipage}
\vfill
\vspace*{-1.5cm}
\begin{minipage}{6cm}
  \begin{picture}(6,6)
    \put(0.,0.){\psfig{file=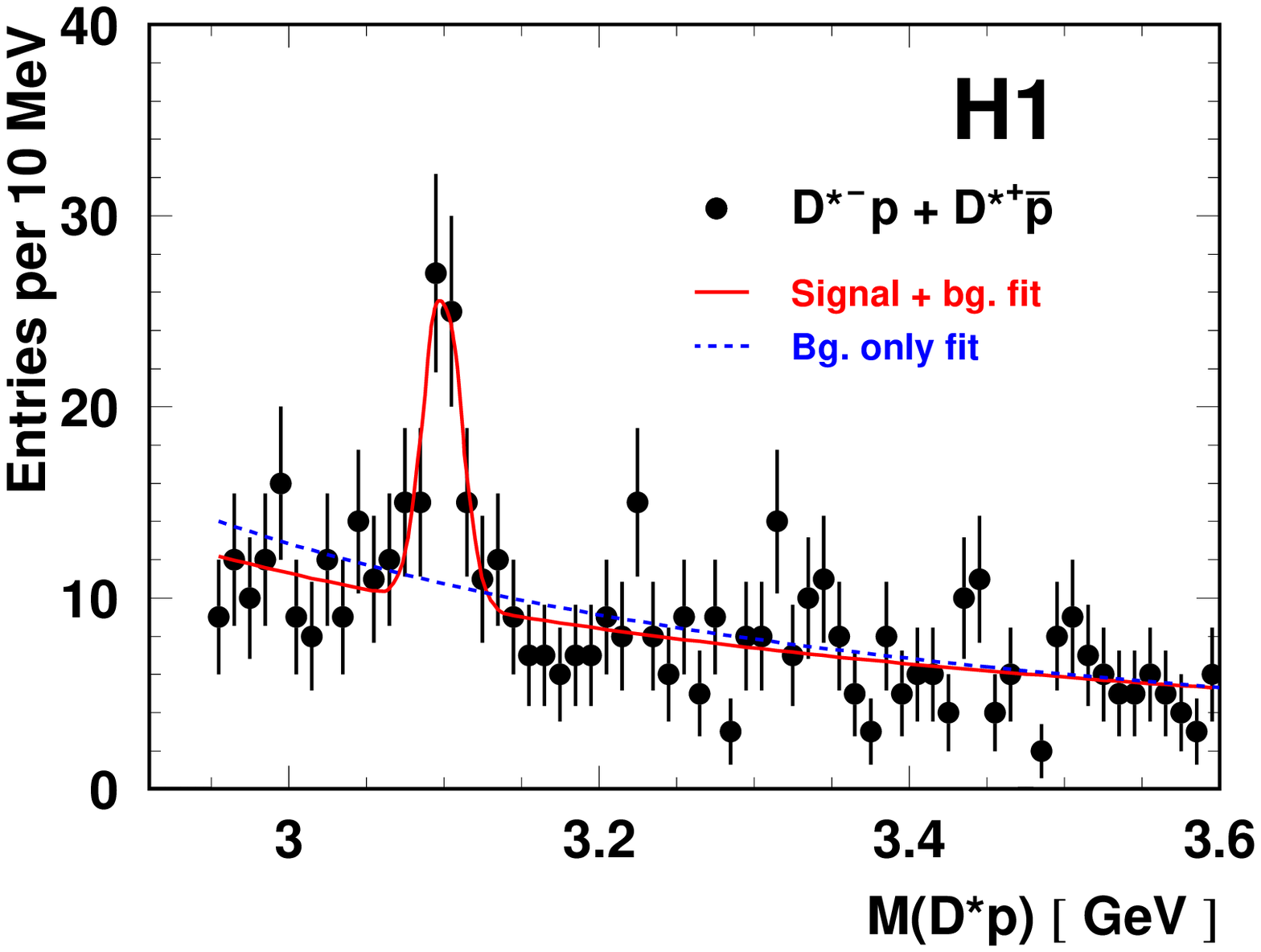,height=4.35cm,clip=}}    
    \put(0.8,4.){\tiny(c)}
  \end{picture}
\end{minipage}
\hfill
\begin{minipage}{6cm}
\begin{picture}(6,6)

  \end{picture}
\end{minipage}
%
\caption{
(a) Invariant mass distribution $M_{D^*p}$ from H1 for $Q^2>1$ GeV$^2$. (b) Momentum distribution of charged particles yielding  $M_{D^*p}$ values falling in the signal and side band regions. (c) $M_{D^*p}$ distribution compared to the fit results with two hypotheses: signal plus background (solid line) and background only (dashed line).
\label{plot_h1Thetac}
}
\end{figure}

The resulting $M_{D^{*-}p}$ distribution in Fig.~\ref{plot_h1Thetac}a shows a clear narrow peak close to the threshold. the data are compared with the absolute expectations from the $D^*$ Monte Carlo (dark histogram) and the non-charmed induced background (light histogram), estimated from the same charge $K\pi$ combinations. No enhancement is seen in any of the background samples. The sum of the two samples reproduces the data well except for the signal region. No peak is observed when selecting either $K\pi\pi_S$ combinations from the $D^*$ side bands or $K\pi$ combinations with masses above the region where the charm contributes or selecting pions instead of protons. The signal is both observed in the $D^{*-}p$ and in the $D^{*+}\overline{p}$ sample with compatible mass, width and rate. No significant enhacement is observed in the like sign $D^* p$ sample. 

In order to check that the signal is due to the decay of a new particle the momentum distribution of the proton candidates without $dE/dx$ cuts has been studied. A background fluctuation should show similar distribution in the signal region and in the side bands. For a real decay, a harder spectrum is expected due to the Lorentz boost of the decaying particle. Fig.~\ref{plot_h1Thetac}b reveals a significantly harder spectrum in the $M(D^*p)$ signal region compared to the side bands. 

The log-likehood fits to the $M(D^*p)$ distribution is shown in Fig.~\ref{plot_h1Thetac}c. The background is parametrised by a power law while a Gaussian is used for the signal. A signal of 51 events is observed with a mass of $3099\pm 3 (stat.) \pm 5 (syst.)$ MeV and a width of $12\pm 3 (stat.)$ MeV consistent with the experimental resolution. The background fluctuation probability has been estimated to be less than $4 \cdot 10^{-8}$.

\begin{figure} [t]
  \unitlength 1cm
\vspace*{0.5cm}
\begin{minipage}{6cm}
  \begin{picture}(6,6)
    \put(0.,0.){\psfig{file=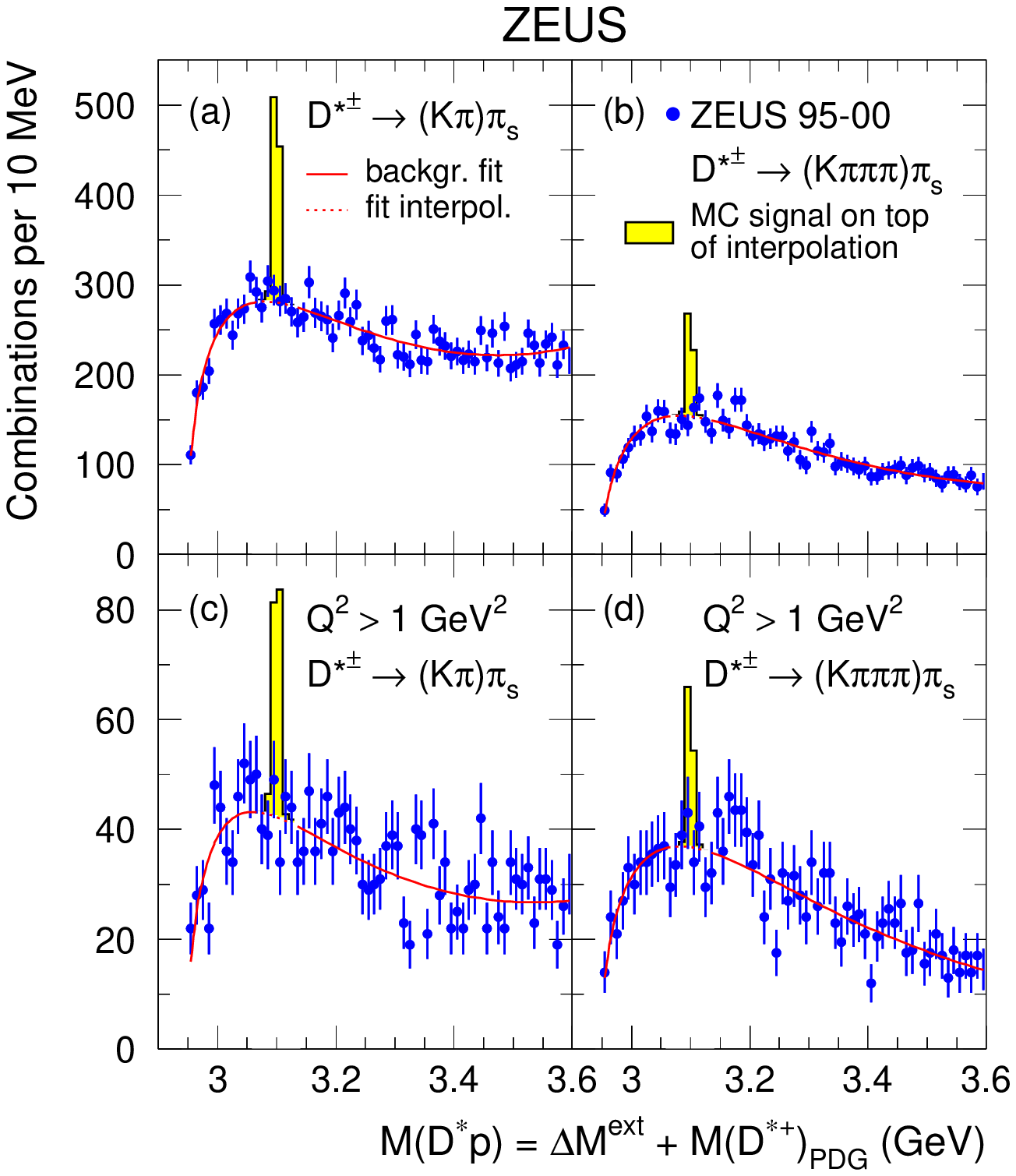,height=7cm,clip=}}    
  \end{picture}
\end{minipage}
\hfill
\begin{minipage}{6cm}
\hspace*{-1.cm}
\begin{picture}(6,6)
    \put(0.,0.){\psfig{file=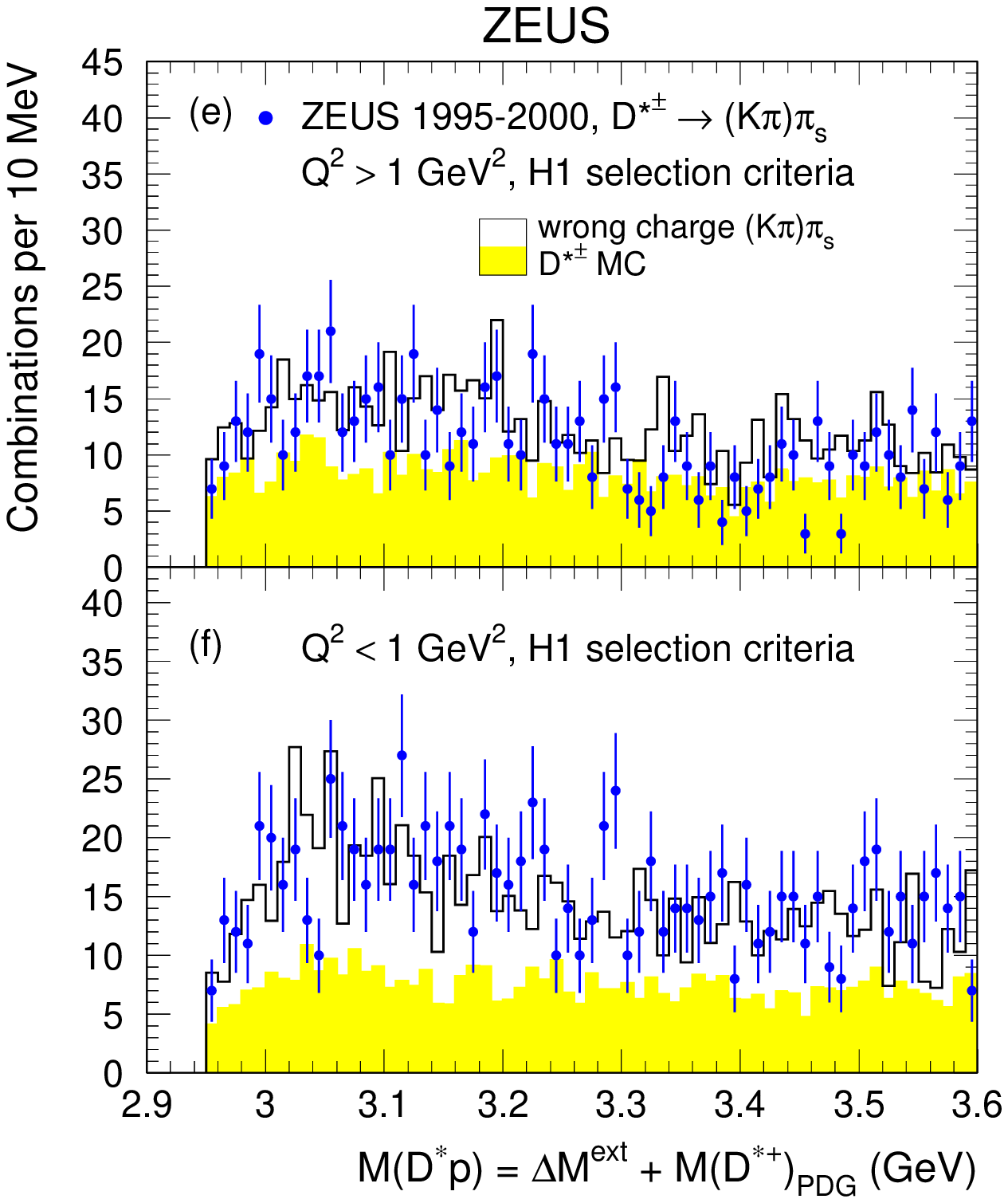,height=7cm,clip=}}
  \end{picture}
\end{minipage}
\caption{
Invariant mass distribution $M_{D^*p}$ from ZEUS for charmed pentaquark candidates selected in the full data sample (a$\&$b) and DIS with $Q^2>1$ GeV$^2$ (c$\&$d), for the $D^*$ decay channels $D^{*\pm}\rightarrow K\pi\pi_S$ (a$\&$c) and $D^{*\pm}\rightarrow K\pi\pi\pi_S$ (b$\&$d).
The solid curves are fits to the background function outside the signal window. The shaded histogram show the Monte Carlo $\Theta_C$ signals, nurmalised to $1\%$ of the number of reconstructed $D^*$ mesons, and added to the fit interpolation (dashed curved) in the signal window. 
Invariant mass distribution $M_{D^*p}$ obtained using H1 selection criteria in (e) DIS with $Q^2>1$ GeV$^2$ and (f) photoproduction with $Q^2<1$ GeV$^2$.
\label{plot_zeusThetac}
}
\end{figure}

A similar search has been performed by ZEUS in both photoproduction and DIS regimes. Data from the years 1995-2000 with an integrated luminosity of 126 pb$^{-1}$ have been analyzed. About 9700 D* candidates are selected for $Q^2>1$ GeV$^2$ and 43000 candidates for all data, and are combined with proton candidates selected via $dE/dx$. The mass distribution $M(D^*p)$ using the same selection criteria as H1 are shown in Fig.~\ref{plot_zeusThetac}(right) for data with $Q^2>1$ GeV$^2$ (Fig.\ref{plot_zeusThetac}e) and $Q^2<1$ GeV$^2$ (Fig.\ref{plot_zeusThetac}f). the data are compared with the absolute expectations from the $D^*$ Monte Carlo (solid histogram) and the combinatorial background (open histogram). No signal is observed at 3.1 GeV. 

Figure~\ref{plot_zeusThetac}(left) shows the mass spectrum $M(D^*p)$ for the full data sample (a and b) and DIS with $Q^2>1$ GeV$^2$ (c and d) obtained with the ZEUS selection criteria~\cite{ZEUSDSTAR}. Two different $D^*$ decay channels $D^{*\pm}\rightarrow (K\pi)\pi_S$ (a and c) and $D^{*\pm}\rightarrow  (K\pi\pi\pi)\pi_S$ (b and d) are considered. No signal is seen in any of the decay channels or kinematic regions considered.  

Upper limits on the fraction of $D^*$ mesons originating from the $\Theta_c^0$ decays, $R=N(\Theta_c \rightarrow D^*p)/N(D^*p)$, were set by ZEUS in the signal window of $3.07 < M(D^*p) < 3.13$ GeV. This window covers the H1 measurement. The $95\%$ confidence level upper limit on the fraction $R$ is $0.23\%$. The upper limit for DIS with $Q^2>1$ GeV$^2$ is $0.35\%$ at $95\%$ C.L.  Thus, the ZEUS results are not compatible with the report of the H1 collaboration of a charmed pentaquark which contributes around $1\%$ of the $D^{*\pm}$ production rate.

\section{CONCLUSIONS}

Recent results from H1, ZEUS, HERMES and HERA-B on searches for exotic baryons in ep collisions, eD scattering and pA scattering at HERA have been presented. ZEUS and HERMES have found evidence for the production of the strange pentaquark $\Theta^{+}$. HERA-B on the contrary has not found any signal compatible with the $\Theta^{+}$ and has obtained limits for its production in  pA scattering.  
ZEUS and HERA-B have not found any evidence for the signal seen by the NA49 collaboration attributed to the $\Xi^{--}$. Both collaborations see a clear signal for the $\Xi^{0}(1530)$ resonance.

H1 has found evidence for the existence of a narrow anti-charmed baryon decaying to $D^{*-}p$. This result has not been confirmed by the ZEUS analysis which has been performed in a similar kinematic region.

Pentaquark searches and studies are still an open issue at HERA. Further studies are needed to understand the positive and negative results obtained in the different searches performed by the four HERA collaborations.


\end{document}